\documentclass[aps,prb,twocolumn,floatfix,showpacs]{revtex4-1}
\usepackage{amsmath,amssymb,graphicx,bm}
\def\veck{\mathbf k}
\def\vecq{\mathbf q}

\begin{document}
\title{Mean-field approximation for thermodynamic and spectral functions of correlated electrons: Strong-coupling and arbitrary band filling}

\author{V\'aclav  Jani\v{s} } 
\affiliation{Institute of Physics, The Czech Academy of Sciences, Na Slovance 2, CZ-18221 Praha  8,  Czech Republic}
\author{Anna Kauch} 
\affiliation{Institute of Physics, The Czech Academy of Sciences, Na Slovance 2, CZ-18221 Praha  8,  Czech Republic}
\affiliation{Institute of Solid State Physics, TU Wien,
Wiedner Hauptstr. 8-10/E138, 1040 Wien, Austria}
\author{Vladislav Pokorn\'y} 
\affiliation{Institute of Physics, The Czech Academy of Sciences, Na Slovance 2, CZ-18221 Praha  8,  Czech Republic}
\email{janis@fzu.cz}

\date{\today}


\begin{abstract}
We present a construction of a mean-field theory for thermodynamic and spectral properties of correlated electrons reliable in the strong-coupling limit. We introduce an effective interaction determined self-consistently from the reduced parquet equations. It is a static local approximation of the two-particle irreducible vertex, the kernel of a potentially singular Bethe-Salpeter equation. The effective interaction enters the Ward identity from which a thermodynamic self-energy, renormalizing the one-electron propagators, is determined. The dynamical Schwinger-Dyson equation with the thermodynamic propagators is then used to calculate the spectral properties. The thermodynamic and  spectral properties of correlated electrons are in this way determined on the same footing and in a consistent manner. Such a mean-field approximation is analytically controllable and free of unphysical behavior and spurious phase transitions.  We apply the construction to the asymmetric Anderson impurity and the Hubbard models in the strong-coupling regime.      
\end{abstract}

\maketitle 

\section{Introduction}

 Tangible electron correlations cause phenomena that remain far from being satisfactorily and fully understood. Only non-perturbative techniques can reliably describe intermediate and strong coupling regimes of correlated electrons. With the increasing computer power various numerically exact methods such as quantum Monte Carlo,\cite{Landau:2014aa,Gull:2011aa} numerical,\cite{Bulla:1998aa,Bulla:2008aa}  density matrix\cite{Schollwock:2005aa,Wolf:2015aa} or functional\cite{Kopietz:2010ab,Karrasch:2008aa} renormalization group, are widely used to obtain quantitative non-perturbative results in the whole range of the interaction strength.  Numerical solutions are, however, restricted to finite-size clusters and relatively small sets of Matsubara frequencies or low-lying excited states. Proximity of the critical points in intermediate and strong coupling demands controlling two-particle and response functions that may diverge in the thermodynamic limit. The singularities in two-particle functions and the critical behavior can be identified and fully controlled only analytically. The singularities and divergencies must be treated separately from the non-divergent quantities so that to reach  stable solutions  and  to avoid instabilities and spurious behavior in the numerical calculations and iterations.      

The simplest analytic static mean-field  approximations work only in the weak coupling and semi-analytic methods must be used to include dynamical fluctuations in the intermediate and strong coupling.  Dynamical mean-field theory\cite{Georges:1996aa} (DMFT), based on the limit of the exact solution in infinite spatial dimensions\cite{Metzner:1989aa} offers an appropriate  framework. Unfortunately, a full solution of the DMFT equations is available only numerically. It hence does not allow us to gain control over the critical behavior (e.g. Mott-Hubbard metal-insulator transition) via analytic functions. Moreover, DMFT addresses directly only one-electron functions, whereby the two-particle ones can be obtained only indirectly with little control of their critical behavior. Extensions of DMFT such as dynamical vertex\cite{Toschi:2007aa} or dual fermions\cite{Rubtsov:2008aa}  partly amend this drawback by addressing also two-particle vertex functions responsible for nonlocal fluctuations. They, however, do not guarantee consistency between the local one-electron and non-local two-electron functions.  

There are two fundamental relations between the one and two-particle many-body Green functions.  The Ward identity between the self-energy and the two-particle irreducible vertex is necessary for making the solution conserving and thermodynamically consistent.\cite{Baym:1961aa,Baym:1962aa}  While the Schwinger-Dyson equation connecting the self-energy with the full two-particle vertex determines the spectral and dynamical properties of the microscopic model.\cite{Schwinger:1951aa,Baym:1962aa,DeDominicis:1964aa} The general unsurmountable problem is that the two equations cannot be simultaneously obeyed  in the approximate treatments.\cite{Janis:1998aa,Janis:2016ac} We suggested in Ref.~\onlinecite{Janis:2016ac} to resolve this inconsistency  by resigning on the full one-particle self-consistency in the Schwinger-Dyson equation used in the standard Baym-Kadanoff construction with the generating Luttinger-Ward functional.  We proposed to use  the two-particle irreducible vertex from a Bethe-Salpeter equation as the generating function from which a thermodynamic self-energy is determined via a (linearized) Ward identity. In this way thermodynamic consistency between the one and two-particle functions is qualitatively achieved. That is, a singularity in the Bethe-Salpeter equation generates the corresponding symmetry breaking in the self-energy. The thermodynamic self-energy does not generally obey the Schwinger-Dyson equation and hence it does not reflect adequately the critical behavior in the spectral function. We thus introduced a spectral self-energy from the Schwinger-Dyson equation in which we used the thermodynamic self-energy to renormalize the one-electron Green function and the two-particle vertex in it.\cite{Janis:2016ac} Although the thermodynamic and spectral self-energies are equal only in the exact solution they generate qualitatively the same critical behavior in thermodynamic and spectral functions. The approximate solutions in this two-particle construction match  the emergence of the order parameter in both self-energies with the unique singularity in the two-particle vertex.     

The existence of two self-energies is a rather common phenomenon of approximate treatments of correlated systems. Apart from approaches that use auxiliary self-energies\cite{Logan:1998aa}  all non-self-consistent or partly self-consistent expansions around mean-field-type solutions introduce two self-energies, the perturbed and unperturbed ones. It is the case when we expand around the static Hartree approximation\cite{Horvatic:1980aa,Kajueter:1996aa,Potthoff:1997aa} or around the local dynamical mean-field solution.\cite{Hettler:2000aa,Maier:2005aa,Toschi:2007aa,Hafermann:2009ab}  The perturbed or corrected self-energy does not change the critical behavior of the unperturbed, mean-field self-energy, but  it may lead to an unphysical behavior such as negative compressibility.  One has to determine the static, spin symmetric part of the physical self-energy via the self-consistency with full propagators and use the Schwinger-Dyson equation with the thermodynamic propagators to determine its dynamical correction. 

The aim of this paper is to extend the two-particle approach introduced in Refs.~\onlinecite{Janis:2007aa} and~\onlinecite{Janis:2016ac} to a full-fledged mean-field theory of correlated electrons at low-temperatures, arbitrary coupling and band filling so that the critical behavior of Bethe-Salpeter equations  is reflected in thermodynamic and spectral functions consistently and on the same footing. The resulting approximation is analytically controllable, free of spurious instabilities and qualitatively correctly describes quantum criticality in  the strong-coupling limit.  

The layout of the paper is as follows. We outline the concept of two self-energies for thermodynamic and spectral properties in Sec. II. A general mean-field approximation with a static local irreducible vertex from the reduced parquet equations is presented in Sec. III.  We apply the approximation on the strong-coupling limit of the single-impurity Anderson model with general local density of states in Sec. IV. The numerical solution of the mean-field approximation for the impurity and lattice models is presented in Sec. V. The last section VI brings conclusions.

\section{Thermodynamic and spectral properties in mean-field approximations}


We demonstrate the construction of mean-field approximations for thermodynamic and spectral properties of strongly correlated electrons on the lattice Hubbard model with a local interaction
\begin{equation}  \label{eq:hh}
  \widehat{H}_H=\sum_{{\bf k}\sigma} \left(\epsilon({\bf k})
    +\mu+\sigma h\right) c^{\dagger}_{{\bf k}\sigma}
  c^{\phantom{\dagger}}_{{\bf k}\sigma}  +
  U\sum_{{\bf i}}\widehat{n}_{{\bf i}\uparrow}\widehat{n}_{{\bf i}
    \downarrow}
\end{equation}
where $h$ is an external magnetic field.  Operators $ c^{\dagger}_{{\bf k}\sigma}$, $c^{\phantom{\dagger}}_{{\bf k}\sigma}$ create and destroy electron with quasi-momentum  $\veck$ and $\widehat{n}_{{\bf i}\sigma}$ is the operator of the electron density on site $\mathbf{R}_{i}$. 

We also use the single-impurity Anderson model (SIAM) with the free conduction electrons hybridized with the interacting local ones $d^{\dagger}_\sigma,d_{\sigma}$
\begin{multline}\label{eq:H-SIAM}
  \widehat{H}_{I} = \sum_{{\bf k}\sigma} \epsilon({\bf k})
  c^{\dagger}_{{\bf k}\sigma} c^{\phantom{\dagger}}_{{\bf k}\sigma}+
  E_d\sum_\sigma d^{\dagger}_\sigma d_\sigma \\ + \sum_{{\bf
      k}\sigma}\left(V^{\phantom{*}}_{{\bf k}}d^{\dagger}_\sigma
    c^{\phantom{\dagger}}_{{\bf k}\sigma} + V^*_{{\bf k}}
    c^{\dagger}_{{\bf k}\sigma} d^{\phantom{\dagger}}_\sigma\right) +
  U\widehat{n}^d_\uparrow\widehat{n}^d_\downarrow\ 
\end{multline}
to demonstrate reliability of the local mean-field approximation as an impurity solver in the strong-coupling, Kondo regime.

\subsection{Thermodynamic and spectral self-energies: Hartree approximation with vertex correction}

Critical behavior of correlated electrons is induced by divergences in response functions and in a Bethe-Salpeter equation in the appropriate two-particle scattering channel. A thermodynamically consistent description demands that the irreducible vertex $\Lambda$, generating a singularity in a Bethe-Salpeter equation, is properly connected with the self-energy so that the latter breaks  the symmetry of the order parameter at the critical point. It is the case when the two functions obey a functional Ward identity.\cite{Baym:1962aa}  In the simplest, Hartree approximation the two-particle irreducible vertex reduces to the bare interaction and the Ward identity
\begin{equation}\label{eq:Ward-Hartree}
\Lambda[U;G] =  \frac{\delta \Sigma[U;G]}{\delta G} = U 
\end{equation}
is exactly solvable.  Here $G$ is the propagator renormalized with the self-energy from the Ward identity.  A solution of Eq.~\eqref{eq:Ward-Hartree} is the  Hartree self-energy   
\begin{equation}\label{eq:SigmaTh-Hartree}
\Sigma_{\sigma}^{T}(k_{n}) =  \frac U{\beta N}\sum_{\veck',n'} G_{-\sigma}(k'_{n'}) \equiv Un^{T}_{-\sigma}
\end{equation}
with the Hartree renormalized propagator   
\begin{equation}\label{eq:GTh-Hartree}
G_{\sigma}(\veck,i\omega_{n}) =  \frac 1{i\omega_{n} + \mu  + \sigma h - \epsilon(\veck) - Un^{T}_{-\sigma}}\,.
\end{equation}
We sometimes use a four-vector notation $k_{n}=(\veck,i\omega_{n})$ for fermionic frequency-momenta and $q_{m}=(\vecq,i\nu_{m})$ for the bosonic ones.

The Hartree self-energy leads to nontrivial response functions and to critical points. It does not, however, obey the Schwinger-Dyson equation with the two-particle vertex derived from the Hartree response functions. The static self-energy serves well for assessing thermodynamic quantities of the equilibrium state but completely fails to account for the dynamical spectral properties. The dynamics of the interacting system is contained in the Schwinger-Dyson equation and hence the latter must be reinstalled in the thermodynamic mean-field theory. 

It is easy to derive the full two-particle vertex from the Hartree magnetic susceptibility. It reads
\begin{equation}
\Gamma_{\uparrow\downarrow}(q_{m}) = \frac U{1 - U^{2}\phi_{\uparrow}(q_{m})\phi_{\downarrow}(q_{m})}\, ,
\end{equation}
where we denoted the electron-hole bubble $\phi_{\sigma}(q_{m}) = (\beta N)^{-1}\sum_{\veck,n}G_{\sigma}(k_{n}) G_{-\sigma}(k_{n} + q_{m})$. 
We use this vertex to introduce another, spectral self-energy from the Schwinger-Dyson equation with this vertex. We take only its dynamical part so that not to break thermodynamic consistency of the static self-energy. That is  
\begin{equation}\label{eq:SigmaSp-Hartree}
\Sigma_{\sigma}^{sp}(k_{n}) = - \frac {U^{2}}{\beta N}\sum_{\vecq,m}\frac{\phi_{\sigma}(q_{m}) G_{-\sigma}(q_{m} + k_{n})}{1 - U^{2}\phi_{\uparrow}(q_{m})\phi_{\downarrow}(q_{m})}\,.
\end{equation} 
We then define the total (physical) self-energy as 
\begin{subequations}\label{eq:GFSigma-full} 
\begin{equation}\label{eq:SigmaT-Full}
\Sigma_{\sigma}(k_{n}) =  \frac U2 \left(n - \sigma m^{T}\right) + \Sigma^{sp}_{\sigma}(k_{n})\,.
\end{equation}
The total particle density $n= n_{\uparrow} + n_{\downarrow}$ in the full self-energy is determined from the one-electron propagator renormalized with the full self-energy 
\begin{equation}\label{eq:GF-full}
\mathcal{G}_{\sigma}(\veck,i\omega_{n}) 
= \frac 1{i\omega_{n} + \mu + \sigma h - \epsilon(\veck) - \Sigma_{\sigma}(\veck,i\omega_{n})}\, ,
\end{equation}
\end{subequations}
while the magnetization $m^{T} = n^{T}_{\uparrow} - n^{T}_{\downarrow}$, Legendre conjugate to the symmetry-breaking field $h$,  is calculated with the thermodynamic propagator, Eq.~\eqref{eq:GTh-Hartree}. This choice of the static part of the physical self-energy is needed not to affect the magnetic instability of the thermodynamic self-energy and to guarantee non-negativity of the physical compressibility calculated from the the full propagator $\mathcal{G}$ in Eq.~\eqref{eq:GFSigma-full}. It means that only the spin-symmetric static part of the physical self-energy is determined self-consistently beyond the thermodynamic self-consistency in propagators $G_{\sigma}(\veck,i\omega_{n})$ in approximate treatments.  

The above construction may serve as a prototype of mean-field approximations for thermodynamic and spectral properties. The Hartree approximation loses reliability in intermediate and strong-coupling regimes. Credible mean-field approximations in the strong coupling must contain a two-particle self-consistency so that to avoid unphysical and spurious poles in the Bethe-Salpeter equations. To improve the Hartree approximation by replacing the static Hartree  self-energy in the Schwinger-Dyson equation by a dynamically renormalized one in the so-called FLEX approximation,\cite{Bickers:1989aa} is a step in the wrong direction. The new renormalized self-energy breaks the Ward identity, shifts the critical point and disconnects the symmetry breaking in the self-energy from the critical point of the two-particle vertex. The correct procedure to go beyond the Hartree approximation is to improve upon the irreducible vertex $\Lambda$ in the Ward identity and to keep two self-energies,  thermodynamic and spectral.

\subsection{General concept of thermodynamic and spectral self-energy}

Thermodynamic consistency of quantum critical behavior demands unambiguous meaning of two-particle Green functions and their singularities. The standard Baym-Kadanoff construction with the Luttinger-Ward generating functional leads to ambiguous definitions of two-particle functions.\cite{Janis:2016ac} We invert the procedure and take the irreducible vertex $\Lambda[U;G]$, the kernel of the singular Bethe-Salpeter equation, as a generating function of thermodynamically consistent theories. The one-electron propagators in the representation of the irreducible vertex are renormalized by a thermodynamic self-energy obeying, to certain degree, the Ward identity.  The Ward identity, being a functional differential equation, cannot be resolved and used to obtain the self-energy for a given irreducible vertex. The full thermodynamic correspondence between the irreducible vertex and the self-energy can, nevertheless, be replaced by a qualitative consistency reached by resolving the Ward identity only linearly with respect to the corresponding  symmetry-breaking field. The latter is the Legendre conjugate to the order parameter induced by the singularity in the Bethe-Salpeter equation. If we assume that the spin-reflection symmetry gets broken at the critical point the Ward identity will be linearized with respect to the external magnetic field $h$.  The thermodynamic self-energy obeying the linearized Ward identity with the given vertex $\Lambda$, similarly to the Hartree approximation, is 
\begin{equation}\label{eq:SigmaT-symbolic}
\Sigma_{\sigma}^{T} =  \Lambda\left[U;G\right]\cdot G_{-\sigma}\,.
\end{equation}
We must symmetrize the  spin-dependent vertex $\Lambda(k_{n},k_{n'};q_{m}) = \left[\Lambda_{\uparrow\downarrow}(k_{n},k_{n'};q_{m}) + \Lambda_{\downarrow\uparrow}(k_{n},k_{n'};q_{m})\right]/2$ so that it depends only on even powers of the symmetry-breaking field. Only with this symmetrization the thermodynamic self-energy $\Sigma_{\sigma}^{T}$ from Eq.~\eqref{eq:SigmaT-symbolic} is a solution of a linearized Ward identity.   The one-particle propagators used in the determination of the two-particle irreducible vertices $\Lambda_{\sigma\sigma'}$ are renormalized with the thermodynamic self-energy 
\begin{multline}\label{eq:G-thermodynamic}
G_{\sigma}(\veck,i\omega_{n}) 
\\
= \frac {1}{i\omega_{n} + \mu + \sigma h -\epsilon(\veck)  - \Sigma^{T}_{\veck,\sigma}(i\omega_{n}) + \Sigma^{T}_{ 0}}\, .
\end{multline}
We introduced a correction $\Sigma_{0}^{T}$ to the thermodynamic self-energy so that to comply with the exact electron-hole symmetry and to keep compressibility non-negative.   The vertex, the self-energy, and the one-electron propagator must obey the following symmetry relations to comply with the electron-hole symmetry $\mu\to U - \mu$
\begin{subequations}\begin{align}
G_{\sigma}(k_{n}) &= - \widetilde{G}_{-\sigma}(-k_{n})  \,,\\
\Sigma^{T}_{\sigma}(k_{n}) &=  2\Sigma_{0}^{T} - \widetilde{\Sigma}^{T}_{-\sigma}(-k_{n}) \,,\\
\Lambda_{\sigma\sigma'}(k_{n},k^{\prime}_{n'};q_{m}) &= \widetilde{\Lambda}_{-\sigma-\sigma'}(-k_{n},-k_{n'};-q_{m}) \,,  
\end{align} \end{subequations}
where the hole functions $\widetilde{X}(i\omega_{n})$ are calculated for $\widetilde{\mu} = U - \mu$. The value of the shift in the self-energy  is $2\Sigma_{ 0}^{T} =\Sigma^{T}_{\uparrow}(\infty) +  \widetilde{\Sigma}^{T}_{\downarrow}(\infty)$ to  assure non-negativity of the compressibility. Here  $\Sigma^{T}_{\sigma}(\infty)$ is the static part of the self-energy at filling $\mu$ and magnetic field $h$. Its explicit form depends on the specific approximation on vertex $\Lambda_{\sigma-\sigma}$. 

The fundamental assumption of this approximation is  that the irreducible vertex $\Lambda$ depends on even powers of the external magnetic field and the thermodynamic self-energy complies with the Ward identity only in the limit of zero magnetic field, where the dependence of vertex $\Lambda$ on the magnetic field can be neglected. The derivative of the thermodynamic self-energy with respect to the magnetic field then diverges at the critical point of the Bethe-Salpeter equation with the irreducible vertex $\Lambda$.  Equation~\eqref{eq:SigmaT-symbolic} determines self-consistently the thermodynamic self-energy  used to renormalize the one-electron propagators $G$ in the perturbation expansion for vertex $\Lambda$. Notice that this construction of the thermodynamically consistent self-energy is similar to  Hedin's GW extension of the Hartree-Fock approximation.\cite{Hedin:1965aa,Aryasetiawan:1997aa,Hedin:1999aa}

The thermodynamic self-energy $\Sigma^{T}$ and the thermodynamic propagators $G$ generally do not guarantee that the critical behavior is correctly described also in the spectral functions. The thermodynamic functions serve well for assessing the static thermodynamic quantities of the equilibrium state. They may, however, give poor results for dynamical spectral properties, in particular within simple (static) mean-field approximations. The dynamical properties of the equilibrium state are extracted from the Schwinger-Dyson equation and the latter must be reinstalled in the approximate theories.  Hence , another, spectral self-energy obeying the dynamical part of the Schwinger-Dyson equation is introduced 
\begin{equation}\label{SDE-symbolic}
\Sigma^{sp}_{\sigma} = -U G_{\sigma}G_{-\sigma}\star\Gamma_{\sigma}\left[U;G\right] \cdot G_{-\sigma} \,.
\end{equation}
The two-particle singlet vertex $\Gamma_{\sigma}\left[U;G\right]$ there is a solution of a Bethe-Salpeter equation 
\begin{multline}
\Gamma_{\sigma}\left[U;G\right]  = \Lambda\left[U;G\right] 
\\
- \Lambda\left[U;G\right] G_{\sigma}G_{-\sigma}\star\Gamma_{\sigma}\left[U;G\right]\, 
\end{multline}
with the same irreducible vertex $\Lambda$ used in the Ward identity. The spectral self-energy has a richer dynamical structure than the thermodynamic one, but shares with it the same critical behavior.  It hence extends the thermodynamic approximation appropriately to the dynamical and spectral functions without affecting the thermodynamic critical behavior. The full physical self-energy is constructed from the spectral one as in the Hartree approximation, Eq.~\eqref{eq:GFSigma-full}. Only the Hartree propagator, Eq.~\eqref{eq:GTh-Hartree} is replaced by the appropriate thermodynamic one, Eq.~\eqref{eq:G-thermodynamic}.

\section{Thermodynamically consistent mean-field approximation}
\label{sec:MFT}

The Hartree mean-field approximation suffers from the deficit that it does not suppress unphysical spurious poles in the two-particle vertex, which makes it fallible in the intermediate coupling and inapplicable in the strong coupling. One must introduce a two-particle self-consistency to prevent the existence of spurious singularities. It is achieved by the parquet scheme. The full parquet approximation\cite{DeDominicis:1962aa,DeDominicis:1964aa} is not analytically solvable and moreover it does not reproduce  the three-peak structure of the DMFT local solution.  It misses the the correct strong-coupling asymptotics and is not a suitable candidate for a mean-field approximation in the strong coupling regime.\cite{Bickers:1989ab,Janis:2006ab} We recently proposed a reduced set of parquet equations that keep the structure of the singularity in the electron-hole scattering channel and balance better multiple electron-hole and electron-electron scatterings than the full set of the parquet equations.\cite{Janis:2016ac} This reduction allows us to reach the critical region of the Bethe-Salpeter equation in the electron-hole channel and the Kondo regime in the impurity model.   

The idea of the reduced parquet equations is to select the singular Bethe-Salpeter equation determining quantum criticality we want to study and decouple the full two-particle vertex to a regular irreducible kernel of the Bethe-Salpeter equation  $\Lambda$  and a reducible diverging part $K$. Another nonsingular Bethe-Salpeter equation is then used to derive a self-consistent, nonlinear equation for the irreducible vertex $\Lambda$. It is important that the self-consistency 
contains only regular expressions and can be solved numerically. The selection of the singular equations depends on the studied problem. It is the electron-hole scattering channel for the repulsive interaction and the electron-electron channel for the attractive coupling. 

\subsection{Effective-interaction approximation}

Since the irreducible vertex $\Lambda$ is regular and bounded we can use a static approximation for it. It means that we neglect all finite fluctuations and keep only large, critical ones in the reducible vertex $K$. If we use the singlet electron-hole singular channel combined with the non-singular electron-electron channel we obtain a simple equation for the irreducible vertex from the reduced parquet equations of Ref.~\onlinecite{Janis:2016ac} 
\begin{equation}\label{eq:Lambda-static}
\Lambda_{\sigma} = \frac U{1 + \Psi_{\sigma}[\Lambda]}
\end{equation}
with
\begin{multline}
\Psi_{\sigma}[\Lambda] = - \frac{\Lambda^{2}}2 \sum_{\alpha=\pm 1}\frac 1\beta\sum_{n} \\
\frac{\phi_{\sigma}(-i\omega_{n} - i\alpha\pi T) G_{\sigma}(i\omega_{n}) G_{-\sigma}(-i\omega_{n})}{1 + \Lambda\phi_{\sigma}(-i\omega_{n} - i\alpha\pi T) }\, 
\end{multline}
and $\Lambda = (\Lambda_{\uparrow} + \Lambda_{\downarrow})/2$.
We extended here the effective-interaction approximation from Ref.~\onlinecite{Janis:2016ac} to non-zero temperatures in that we added the average over the lowest fermionic Matsubara frequencies to  the vertex functions and kept the irreducible vertex a real constant. Such a straightforward extension allows us to represent the screening factor analytically with  Fermi, $f(\omega)=1/[\exp(\beta \omega) + 1]$,  and Bose, $b(\omega)=1/[\exp(\beta \omega) - 1]$, distribution functions
\begin{multline}
\Psi_{\sigma}[\Lambda] = \frac{\Lambda^{2}}{\pi}\int_{-\infty}^{\infty}d\omega \left\{f(\omega) \Re\left[\frac{\phi_{\sigma}(-\omega -i\pi T)}{1 + \Lambda\phi_{\sigma}(-\omega -i\pi T)}\right] 
\right. \\ \left. 
\times \Im\left[G_{\sigma}(\omega)G^{*}(-\omega)  \right] %
- b(\omega)\Im \left[\frac{\phi_{\sigma}^{*}(-\omega )}{1 + \Lambda\phi_{\sigma}^{*}(-\omega)}\right]
\right. \\ \left. 
\times  \Re\left[G_{\sigma}(\omega - i\pi T)G(-\omega + i\pi T)  \right] \right\} \, 
\end{multline}
and 
\begin{multline}
\phi_{\sigma}(\omega)
 = -  \int_{-\infty}^{\infty} \frac{dx}{\pi } f(\omega) \left[G_{-\sigma}(x + \omega ) \Im G_{\sigma}(x)
 \right. \\ \left. 
 +\  G_{\sigma}^{*}(x - \omega)\Im G_{-\sigma}(x) \right]  \, . 
\end{multline}  
The variables along the cuts are always taken as the limit from the upper complex half-plane, that is $G(x) = G(x + i0^{+})$ and $G^{*}(x) = G(x - i0^{+})$ denotes complex conjugation.    

The full singlet local electron-hole vertex reads
\begin{equation}
\Gamma_{\sigma}(\omega) = \frac{\Lambda}{1 + \Lambda \phi_{\sigma}(\omega)}\, . 
\end{equation}

\subsection{Thermodynamic and spectral self-energy and band structure}
\label{sec:Selfenergies}

The two self-energies used in this mean-field approximation lead to two particle occupations. The thermodynamic self-energy renormalizes the bare propagator to a thermodynamic one the local element of which is
\begin{equation}
G_{\sigma}(\omega) = \int_{-\infty}^{\infty}\frac{d\epsilon \rho(\epsilon)}{\omega + i0^{+} + \bar{\mu}_{\sigma} -\epsilon} \,  
\end{equation}
where 
\begin{equation}
\bar{\mu}_{\sigma} = \mu + \sigma h - \frac{U - \Lambda}2  - \Lambda n^{T}_{\sigma} \,.  
\end{equation}
The thermodynamic occupation number  then is
\begin{equation}\label{eq:nT}
n^{T}_{\sigma}= \int_{-\infty}^{\infty}d\epsilon \rho(\epsilon)f(\epsilon -\bar{\mu}_{\sigma}) \,.
\end{equation}
The effective chemical potential $\bar{\mu}$ in the thermodynamic Green function is determined from a doping parameter $x=\mu - U/2$ at zero magnetic field and temperature as
 \begin{equation}\label{eq:barmu-x}
x= \bar{\mu} +  \Lambda\ \text{sign}(\bar{\mu})\int_{0}^{\lvert\bar{\mu}\rvert}d\epsilon \rho(\epsilon) \,.
\end{equation} 
The physical particle density is determined from the propagator renormalized with the full self-energy, Eq.~\eqref{eq:GFSigma-full}, 
\begin{multline}
n_{\sigma} = \int_{-\infty}^{\infty}\frac {d\omega}{\pi} f(\omega)\int_{-\infty}^{\infty}d \epsilon
\\
\frac {\rho(\epsilon)}{\omega + i0^{+} + \mu + \sigma h - \epsilon - \frac U2 \left(n - \sigma m^{T}\right) - \Sigma^{sp}_{\sigma}(k_{n})}\, .
\end{multline}
The spectral self-energy of the effective-interaction approximation has an explicit representation
\begin{multline}\label{eq:Sigma-spectral}
\Sigma_{\sigma}^{sp}(\omega)
\\
 = -\frac{U\Lambda}{\pi} \int_{-\infty}^{\infty} dx \left\{b(x) G_{-\sigma}(x + \omega)\Im\left[\frac{\phi_{\sigma}(x)}{1 + \Lambda\phi_{\sigma}(x)} \right]  
\right. \\ \left.
  - f(x) \frac{\phi^{*}_{\sigma}(x - \omega)}{1 + \Lambda \phi^{*}_{\sigma}(x - \omega)} \Im G_{-\sigma}(x)\right\}\,. 
\end{multline}

The thermodynamic and spectral Green functions are not the same in this approximate treatment, hence they lead to different band fillings and band edges for the given input parameters, density of states, chemical potential and interaction strength.   The band edges of the thermodynamic propagator are at $\lvert\bar{\mu}\rvert = w$, if the bare band has support on interval $(-w,w)$. Then $\Im\phi(\omega) =0$ and consequently also $\Im\Sigma(\omega) =0$ for $\lvert\bar{\mu}\rvert = w$. The effective interaction equals the bare one, $\Lambda = U$ at the band edges.  The lower band edge is  $\mu_{l} = - w  + (U - \Lambda)/2$, while the upper band edge is $\mu_{u} =  w + (U - \Lambda)/2 + U$.  The band edges of the spectral function will have the same structure, only the shift of the center of the band is different. We find $w_{l} = -w + U/2 - \Sigma_{0} = -w$ and $w_{u} = w + U$.  The bandwidth is the same, only the center of the band of the thermodynamic Green function is shifted with respect to the exact value reproduced by the spectral self-energy.

\subsection{Magnetic susceptibility}

The critical behavior of the thermodynamic system is experimentally determined from response functions. It is a transition to a magnetic order in the electron systems with the repulsive interaction. The thermodynamic information about the magnetic critical behavior is contained in the (dynamical) magnetic susceptibility,  $\chi(\vecq,i\nu_{m})$. The static susceptibility diverges at the transition point at a specific transfer momentum $\vecq$, dependent on which type of magnetic order emerges at low temperatures. Our approximate theory based on the reduced parquet equations generates two self-energies and two one-electron Green functions. They then lead to two susceptibilities but with qualitatively the same critical behavior. It means that they both determine the same critical point of the transition and the long-range order emerges at the same point of the phase diagram. We evaluate explicitly both susceptibilities and compare their values in numerical calculations.   

To determine the susceptibility from the Green function using the thermodynamic self-energy, that is constant and  renormalizes the chemical potential to $\bar{\mu}$, we will need a momentum-dependent electron-hole bubble  defined in the Matsubara formalism  
\begin{subequations}\begin{multline}
\phi(\vecq, i\nu_{m}) \\
= \frac 1\beta\sum_{n} \frac 1N\sum_{\mathbf{k}}G(\veck,i\omega_{n}) G(\veck + \vecq,i\omega_{n} + i\nu_{m}) \,.
\end{multline}
The sum over Matsubara frequencies can easily be analytically continued  and we obtain an analytic representation of the dynamical bubble
\begin{align}
\phi(\mathbf{q},\omega) &=  \frac 1N\sum_{\mathbf{k}} \frac {f(\epsilon(\mathbf{k}) - \bar{\mu}) - f(\epsilon(\mathbf{k} + \mathbf{q}) - \bar{\mu})}{\omega - \epsilon(\mathbf{k} + \mathbf{q}) + \epsilon(\mathbf{k})}\,. 
\end{align}\end{subequations}

In case of the ferromagnetic response with $\vecq = \mathbf{0}$ and $\epsilon(\veck + \vecq) = \epsilon(\veck)$ we have
\begin{equation} 
\phi_{F}(\mathbf{0},0) = \int_{-\infty}^{\infty} d \epsilon \rho(\epsilon)  f^{\prime}(\epsilon - \bar{\mu}) \, .  
\end{equation}
It reduces at zero temperature to $\phi(\mathbf{0},0) = -\rho(\bar{\mu})$ and we obtain a renormalized Stoner criterion for the ferromagnetic instability $1 = \Lambda \rho(\bar{\mu})$. In case of the antiferromagnetic response, $\vecq = \mathbf{Q}$, where vector $\mathbf{Q}$ is defined as  $\epsilon(\veck + \mathbf{Q}) = - \epsilon(\veck)$, we obtain  
\begin{equation} 
\phi_{AF}(\mathbf{Q},0) = \mathcal{P} \int_{-\infty}^{\infty} d \epsilon f(\epsilon - \bar{\mu}) \frac{\rho(\epsilon)}{\epsilon} \, .  
\end{equation}

The dynamical thermodynamic susceptibility then is
\begin{equation}\label{eq:chiT-dynamic}
\chi^{T}(\mathbf{q}, \omega) = - \frac {2\phi(\mathbf{q},\omega)}{1 + \Lambda \phi(\mathbf{q},\omega)} \,.
\end{equation}
The physical dynamical susceptibility has a more complex expression and it mixes the thermodynamic Green function $G(\veck,i\omega_{n})$ and the full one $\mathcal{G}(\veck,i\omega_{n})$ in a representation
\begin{multline}\label{eq:chi-dynamic}
\chi(\vecq, i\nu_{m}) = - \frac{2}{1 + \Lambda\phi(\mathbf{q},i\nu_{m})} \frac 1\beta \sum_{n}\frac{1}{N}\sum_{\mathbf{k}} \mathcal{G}(\mathbf{k},i\omega_{n})  \phantom{\frac12}
\\
\times \mathcal{G}(\mathbf{k} + \mathbf{q},i\omega_{n} + i\nu_{m}) \left[1 + \left(\Lambda - U\right) \phi(\mathbf{q},i\nu_{m}) 
\right. \\ \left. 
+  \frac 1\beta \sum_{m'}\frac {U\Lambda \phi(i\nu_{m'})}{1 + \Lambda\phi(i\nu_{m'})}
\frac{1}{N}\sum_{\mathbf{k'}} G(\mathbf{k}',i\omega_{n + m'})
\right. \\ \left. 
\phantom{\frac12}\times  G(\mathbf{k}' + \mathbf{q} ,i\omega_{n + m'} + i\nu_{m}) \right] \,.  
\end{multline}
It is evident that both the thermodynamic and the physical  susceptibilities share the same critical behavior at $1 + \Lambda \phi(\vecq,0) = 0$. Since the spectral self-energy is frequency dependent we cannot sum over the Matsubara frequencies explicitly in the physical susceptibility. It is easy to verify that at half filling ($\bar{\mu}=0$) the system is antiferromagnetically ordered at zero temperature for arbitrary interaction strength $U>0$.  The electron-hole bubble  logarithmically diverges $\phi_{AF}(\mathbf{Q},0) =  \rho(0)\ln(\lvert\bar{\mu}\rvert /w)$ for $\rho(\epsilon)= \rho(-\epsilon)$, $T=0$ and for $\bar{\mu}\to0$. As other static mean-field solutions, the effective-interaction approximation becomes unreliable in the description of the lattice models in low-spatial dimensions, $d=1,2$. A dynamical and non-local extension of the static effective interaction is needed to improve the reliability to low-dimensional systems.

\section{Single-impurity Anderson model: Strong-coupling limit}

The effective-interaction approximation was introduced to describe qualitatively correctly quantum critical behavior and quantum phase transitions in the intermediate and strong coupling. One can use the single-impurity Anderson model  to demonstrate reliability of this mean-field approximation in the strong-coupling regime. There is no critical point in SIAM for finite interaction strength, but the Kondo regime with an exponentially small Kondo temperature is a critical region of the singlet electron-hole vertex. We demonstrated in Ref.~\onlinecite{Janis:2016ac} that both the spectral function and the magnetic susceptibility at zero temperature and in the charge and spin symmetric state lead to a qualitatively correct Kondo temperature, the logarithm of which linearly depends on the bare interaction strength.  Here we apply the effective interaction to describe the strong-coupling limit of the spin-symmetric equilibrium state of SIAM away from half filling and for arbitrary density of states of the localized electrons.     

The exact solution of the impurity model is known only for  the Lorentzian density of states (DOS),
\begin{equation}\label{eq:DOS-Lorentz}
\rho(\omega) = \frac 1\pi \frac{\Delta}{\omega^{2} + \Delta^{2}}\,,
\end{equation} 
where $\Delta = \pi\sum_{\veck}\left\lvert V_{\veck}\right\rvert^{2}\delta (E_{F} -\epsilon(\veck)$) is the effective hybridization between the local and conduction electrons at the Fermi energy and sets the energy unit. Our mean-field approximation allows us to formulate the strong-coupling limit for arbitrary form of the density of states of the local electrons. The only difference to the lattice models is that the impurity model does not allow for a magnetic order and the equilibrium state remains the local Fermi liquid to infinite interaction strength. Our approximate thermodynamic and spectral self-energies for the impurity model are the mean-field ones from Sec.~\ref{sec:Selfenergies}. We apply the approximation to the zero-temperature spin-symmetric state, that is $T=0,h=0$.  

The strong coupling limit $U\to \infty$ generates an exponentially small Kondo scale for the charge-symmetric situation, $x=0$. The Kondo scale emerges there due to the sum rule making the density of states at the Fermi level independent of the interaction strength. It is no longer true if we move away from the charge-symmetric case. The strong-coupling limit then depends on the way we reach the limit of infinite interaction strength. 

In the first step we fix the effective chemical potential $\bar{\mu}$ being equivalent to keeping the thermodynamic occupation number $n^{T}$ constant. The strong-coupling limit then is close in behavior to that from half-filled case with a critical behavior in the two-particle vertex $\Gamma(\omega)$. Its critical behavior for a dimensionless Kondo scale $a = 1 + \Lambda \phi(0)\ll 1$ is    
\begin{subequations}
\begin{align} 
\phi'(0) &= - \pi \rho(\bar{\mu})\,, \\
\Gamma(\omega) &\doteq \frac {\Lambda}{a - i\pi\Lambda \rho(\bar{\mu})^{2}\omega}\,,
\end{align}
\begin{multline}
\phi(0) =  - 2\int_{-\infty}^{0}d x \rho(\bar{\mu} + x)
\\
\times \int_{0}^{\infty} d \epsilon\frac{\rho(\bar{\mu} + x +\epsilon) - \rho(\bar{\mu} + x - \epsilon)}{\epsilon}\,.
\end{multline}
\end{subequations}
The screening of the bare interaction and the Kondo asymptotic scale read
\begin{subequations}
\begin{align} 
\psi &\doteq \frac{\lvert \phi(0)\rvert \lvert G(0)\rvert^{2}\Lambda}{\pi^{2}\rho(\bar{\mu})^{2}} \lvert \ln a\rvert \,, \\
a &= \exp\left\{ - \frac{U \lvert \phi(0)\rvert }{1 + \lvert \Re G(0)\rvert^{2}/\lvert \Im G(0)\rvert^{2}}\right\} \,.
\end{align}
where
\begin{align} 
\Im G(\omega) &= - \pi \rho(\bar{\mu} + \omega)\,, \\
\Re G(\omega)  &= -2\int_{0}^{\infty} d \epsilon\frac{\rho(\bar{\mu} + \omega +\epsilon) - \rho(\bar{\mu} + \omega - \epsilon)}{\epsilon} \,.
\end{align}
\end{subequations}
For small dopings and the Lorentzian density of states  $\lvert\bar{\mu}\rvert\ll \Delta$ we obtain $2\bar{\mu} \doteq x$ and
\begin{align}
a(x) & \doteq \exp\left\{- \frac{U}{\pi \Delta}\left( 1 - \frac {x^{2}}{2\Delta^{2}}\right) \right\} \,, 
\end{align}
while the exact Kondo scale determined from the magnetic susceptibility of the Bethe ansatz solution is $a = \exp\{-\pi (U^{2}/4 - x^{2})/2\Delta U \}$.\cite{Tsvelick:1983aa,Hewson:1993aa} The approximate solution correctly reproduces the quadratic increase of the Kondo scale with parameter $x \ll \Delta^{2}$ but misses the exact prefactor. The Kondo regime in the approximate solution sets in for a much stronger interaction strength.  While the Kondo regime sets in the exact solution for $U\gg 8(\Delta + x^{2}/2U)/\pi$ in the approximate solution it is  for  
\begin{equation}\label{eq:U-barmu}
 \frac{U}{\pi \Delta} \gg  \left(1 +  \frac{\bar{\mu}^{2}}{\Delta^{2}}\right)^{2} \approx 1 + \frac{x^{2}}{2\Delta^{2}}\,.  
\end{equation}
A worse reliability of the effective interaction away from half filling stems from the suppressed dynamics of multiple scatterings from the electron-electron channel that contribute to vertex $\Lambda$. This non-divergent vertex was replaced by a constant. Fluctuations due to multiple electron-electron scatterings, although non-divergent,  become more relevant the farther we are from half filling and affect the way the critical, strong-coupling Kondo regime is reached. A more elaborate approximation with both two-particle vertices dynamical is needed to reach better results in the strong-coupling regime away from half filling.

The two-particle vertices are further used to determine the spin susceptibility and its dependence on the dimensionless Kondo scale $a$. The physical susceptibility at zero magnetic field can be represented within the effective-interaction approximation as\cite{Janis:2016ac}
\begin{equation}\label{eq:chi-zero}
\chi = - \frac 2\beta \sum_{n}\mathcal{G}(i\omega_{n})^{2}\left[ 1 - \frac{ UX(i\omega_{n})}{a}\right] \, ,
\end{equation}
where we introduced a two-particle  function
\begin{multline}\label{eq:Xn-def}
X(i\omega_{n}) = \frac 1\beta\sum_{m}\frac{G(i\omega_{n} + i\nu_{m})}{\left[1 + \Lambda\phi(i\nu_{m})\right]^{2}}\left\{ G(i\omega_{n} + i\nu_{m})
\right. \\ \left.
\times\left[1 + \Lambda\phi(i\nu_{m})\right] + \Lambda\left[\kappa(i\nu_{m}) - \kappa(-i\nu_{m}) \right]\right\}\,,
\end{multline}
in which 
\begin{equation}
\kappa(i\nu_{m})=  \frac 1\beta \sum_{n} G(i\omega_{n} + i\nu_{m}) G(i\omega_{n})^{2}\,.
\end{equation}

The Kondo scale is a two-particle quantity that regularizes the pole of the critical two-particle vertex. It enters the one-particle spectral function via the Schwinger-Dyson equation matching the two-particle vertex and the one-particle spectral self-energy, Eq.~\eqref{eq:Sigma-spectral}. The dynamical part of the spectral self-energy in the Kondo limit $a \ll 1$ with the dominant logarithmic singularity in the effective-interaction approximation reads
 \begin{subequations}\label{eq:SE-asymptotic}
 \begin{align}\label{eq:SE-ReS}
   \Re\Sigma^{sp}(\omega) &\doteq \frac {U }{\Lambda \pi^2
     \rho_0^2}\bigg[ |\ln a|\ \Re G(\omega) \nonumber \\ &\qquad +
   \arctan\left(\frac{\Lambda\pi\rho_0^2\ \omega}{a}\right) \Im
   G(\omega)\bigg] \ , 
   \\ \label{eq:SE-ImS} 
   \Im\Sigma^{sp}(\omega) &
   \doteq \frac {U}{2\Lambda \pi^2 \rho_0^2}\ln \left[ 1 + \frac
     {\Lambda^2 \pi^2\rho_0^4\ \omega^2} {a^2}\right] \Im
   G(\omega)\ .
 \end{align}
 \end{subequations}
The spectral self-energy is significantly frequency modulated only on a Kondo length scale of order $a$. Consequently the spectral function develops a quasiparticle peak near the Fermi energy of the same width.  

The strong-coupling limit for the fixed doping parameter $x$ is due to Eq.~\eqref{eq:barmu-x} qualitatively the same as that of the fixed thermodynamic-occupation number $n^{T}$. Quite a different strong-coupling limit is obtained if we fix the charge density $n$. When $U\to\infty$ at a fixed $n$, 
\begin{equation}
\bar{\mu} + \frac \Lambda\pi \arctan\left(\frac{\bar{\mu}}\Delta \right) \propto \frac U2(n - 1)
\end{equation} 
and the effective chemical potential $\bar{\mu}$ goes to $\pm \infty$ linearly with $U$ according to whether $n>1$ or $n<1$. The Kondo regime cannot then be reached, since Eq.~\eqref{eq:U-barmu} cannot be satisfied. There is hence no Kondo regime in the strong-coupling limit for the fixed particle density $n\neq 1$. The two-particle vertex $\Gamma(\omega)$ reaches maximum for $\omega=0$ at a finite interaction strength $U_{m}(n)\to \infty$ for $n\to 1$.

\section{Numerical results}

We use the effective-interaction approximation for the irreducible vertex $\Lambda$  from Eq.~\eqref{eq:Lambda-static} at zero temperature to calculate mean-field spectral and response functions in the strong-coupling limit of the Fermi-liquid phase of the Anderson impurity and Hubbard models at zero temperature and away from half filling.  Numerical results for non-zero temperatures will be presented elsewhere.  

\subsection{Impurity model}
\label{sec:IMN}

\begin{figure}
\includegraphics[width=8.5cm]{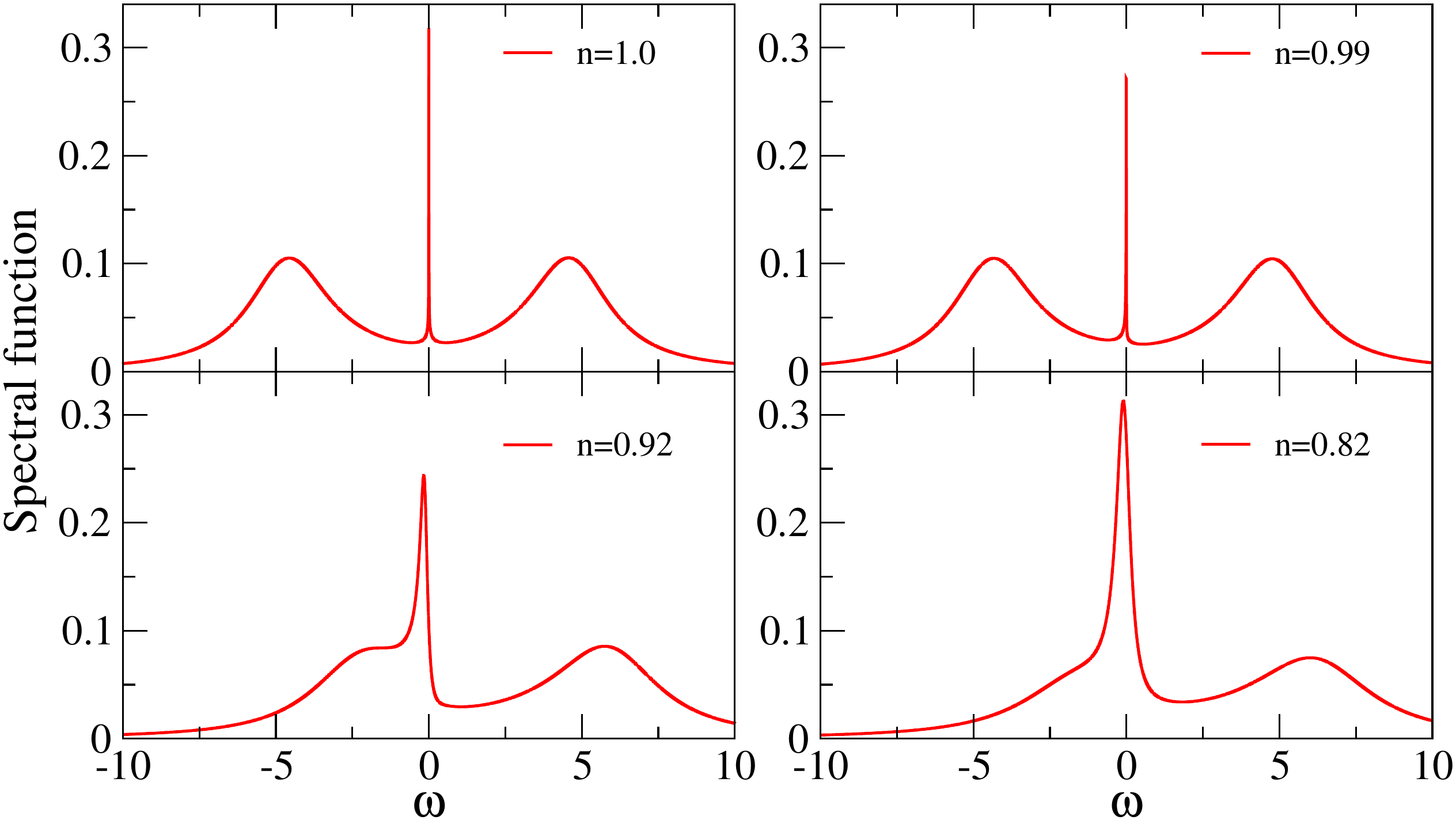}
\caption{(Color online)  Spectral function of SIAM at zero temperature and $U=12$. We used Lorentzian density of states with $\Delta$ taken as the energy unit. Doping the state with holes leads to  merging of the central quasiparticle peak with the lower Hubbard satellite band.   \label{fig:spectral_sc}}
\end{figure}

\begin{figure}
\includegraphics[width=8cm]{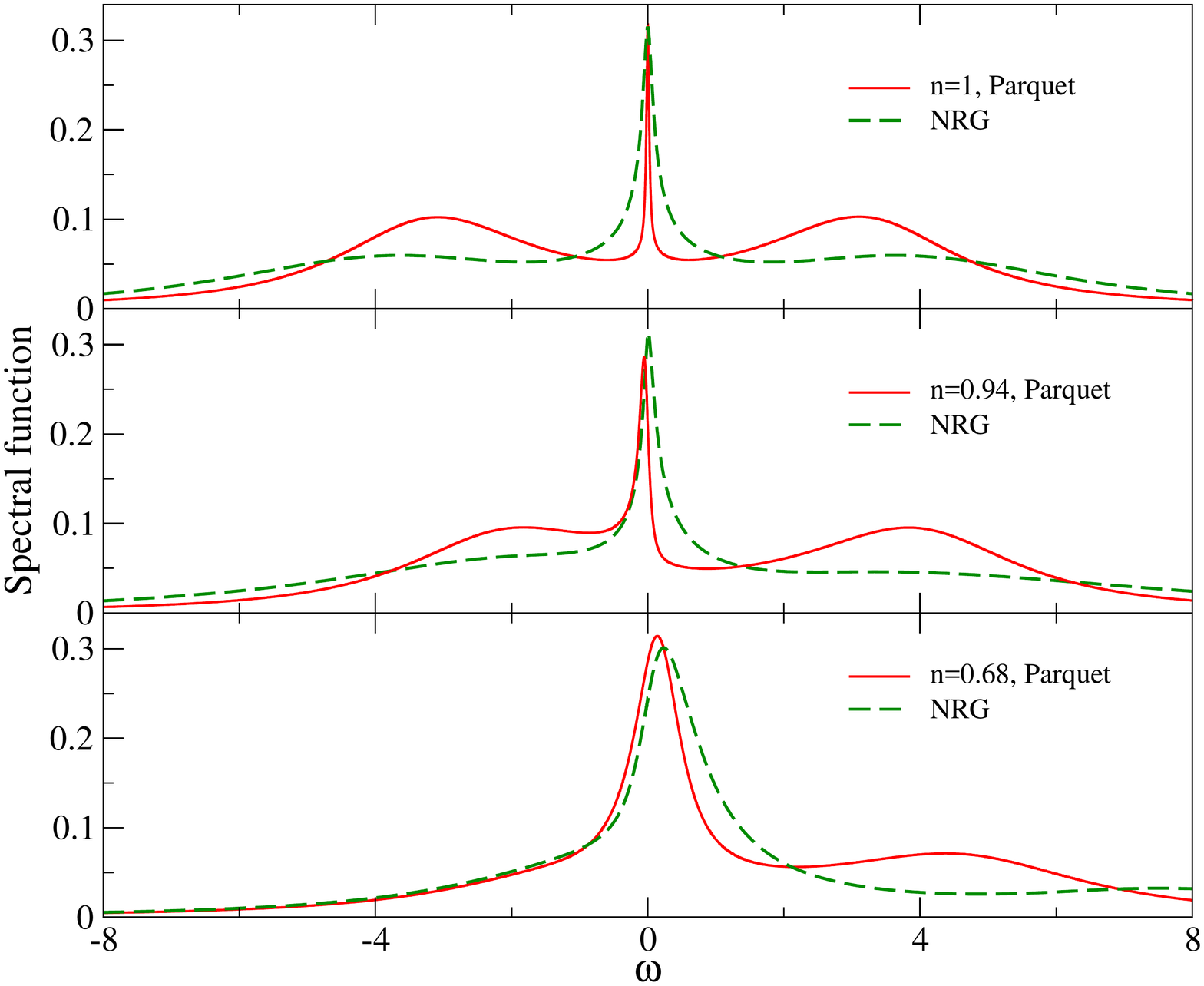}
\caption{(Color online)  Spectral function of SIAM for $U=8$ compared with NRG result.    \label{fig:spectral_NRG}}
\end{figure}

We first evaluate the spectral function calculated from Eq.~\eqref{eq:GF-full} with the Lorentzian density of states, Eq.~\eqref{eq:DOS-Lorentz}. Figure~\ref{fig:spectral_sc} shows how the spectral function develops when doping the electron-hole symmetric solution, $n = n_{\uparrow} + n_{\downarrow}=1$ with holes. We can see that the height of the central quasiparticle peak decreases and the lower Hubbard satellite approaches the Fermi surface. Finally, the lower Hubbard band merges with the central one. We compared the spectral function calculated within the effective-interaction approximation with the spectral function from the numerical renormalization group (NRG) for $U=8$ in Fig.~\ref{fig:spectral_NRG}. The results were obtained by using the  NRG Ljubljana Code.\cite{NRGLjubljana}  The central quasiparticle peak of the reduced parquet equations is narrower than that of NRG, since the logarithmic corrections of the exact solution are missing in our approximation. The mean-field approximation reproduces, however,  the correct strong-coupling asymptotics with the logarithm of the Kondo scale linearly dependent on the bare interaction for the fixed effective chemical potential $\bar{\mu}$.  Moreover, the Hubbard satellites are better reproduced in the high-energy region, where NRG gives less accurate results. The  overall tendency of the spectral function where the lower Hubbard band absorbs the quasiparticle peak is in both solutions the same. 

\begin{figure}
\includegraphics[width=8cm]{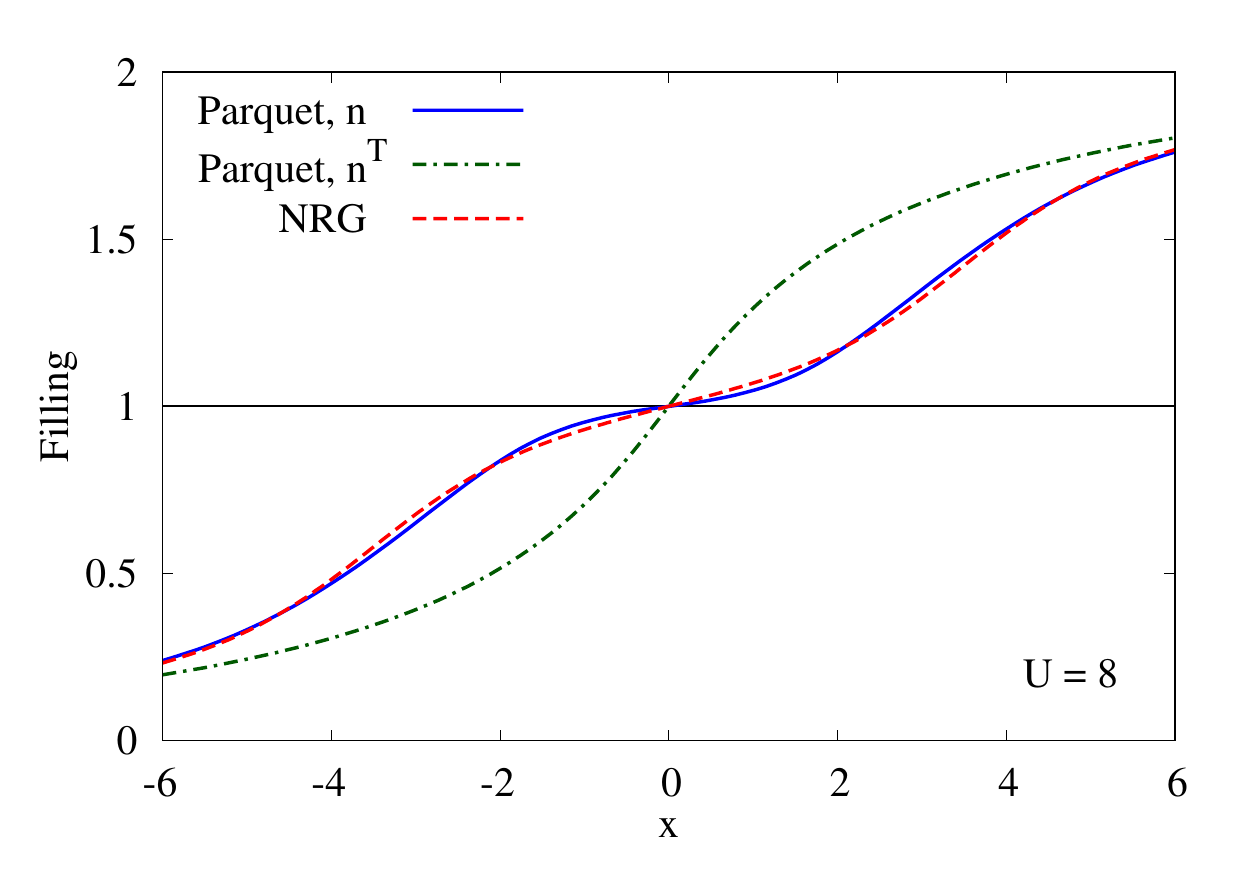}
\caption{(Color online) Band fillings $n$  and $n^{T}$ as a function of the doping parameter $x=-E_{d} -U/2$ from the static local approximation in the reduced parquet equations compared with the NRG result.   \label{fig:spectral_n}}
\end{figure}

We plotted in Fig.~\ref{fig:spectral_n} the dependence of the auxiliary thermodynamic occupation number $n^{T}$ from Eq.~\eqref{eq:nT} and the real (physical) occupation $n$, self-consistently determined from Eq.~\eqref{eq:GFSigma-full}, on the doping parameter $x=-E_{d} -U/2$ and compared them with the charge density from the numerical renormalization group. The physical occupation number quite well fits the NRG data for all band fillings. The thermodynamic occupation number $n^{T}$ is off in the similar way as the Hartree approximation. Both approximate solutions $n(x)$ and $n^{T}(x)$ generate nonnegative compressibility. It is the consequence of the self-consistency with which they are determined.  

\begin{figure}
\includegraphics[width=8cm]{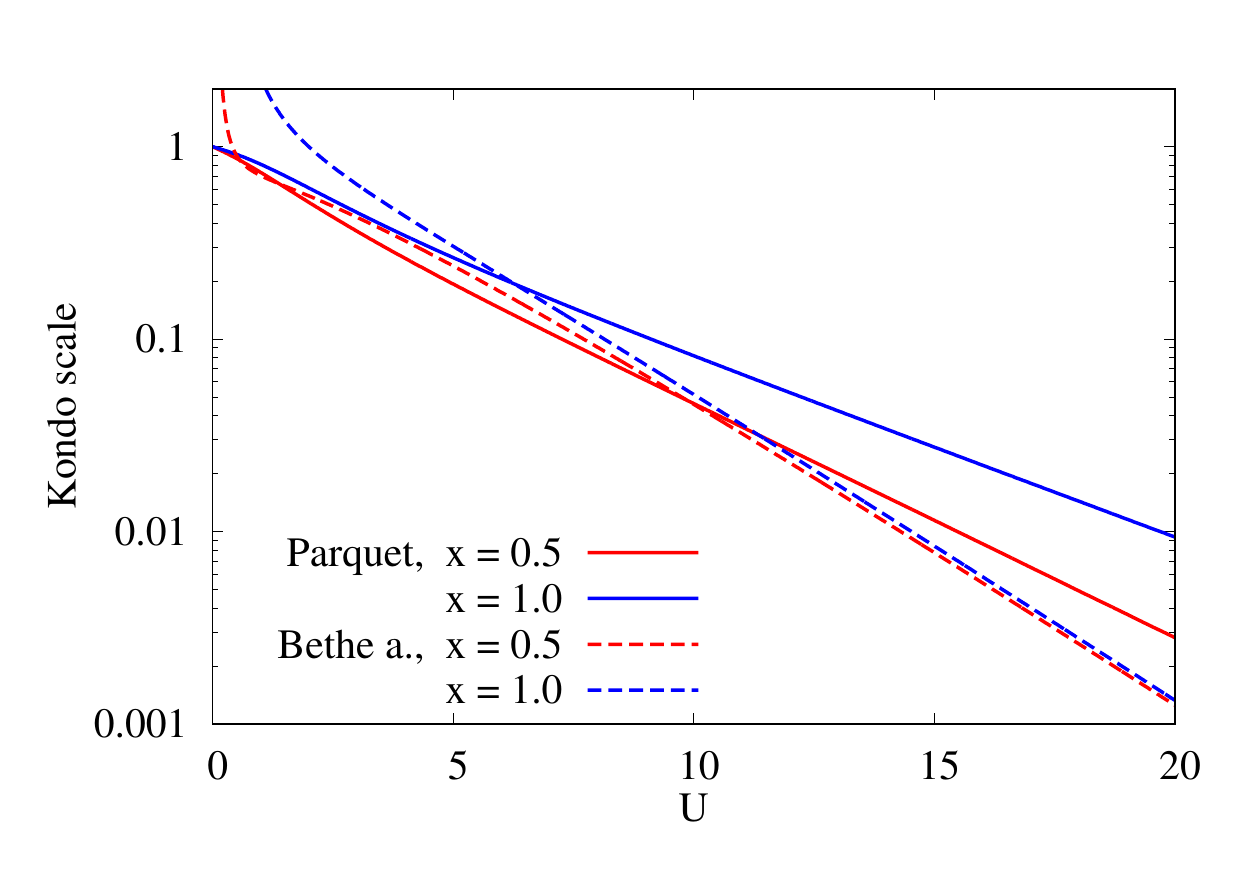}
\caption{(Color online) Kondo scale calculated from the denominator of the vertex function $K(\omega)$ at the Fermi energy $\omega=0$ as a function of the interaction strength $U$ compared for two doping parameters $x$ with the exact Bethe ansatz result.   \label{fig:Kondo_U}}
\end{figure}

The mean-field approximation allows us to evaluate the strong-coupling limit and to check the behavior of the Kondo scale there away from half filling. The reduced parquet equations in the effective-interaction approximation reproduce the Kondo scale only in the critical region of the metal-insulator transition where it is exponentially small. We can expect that the farther from half filling the worse the Kondo scale from the Bethe ansatz will be reproduced.  We compared in Fig.~\ref{fig:Kondo_U} the Kondo scale dependence on the interaction strength for the fixed doping parameter with the Bethe ansatz solution. Although the numerical values of the Kondo scale differ, the linearity of the logarithm of the Kondo scale is correctly reproduced. The shift in the numerical values is caused by a difference in the nonuniversal prefactors that the parquet approximation does not reproduce. 

\begin{figure}
\includegraphics[width=8cm]{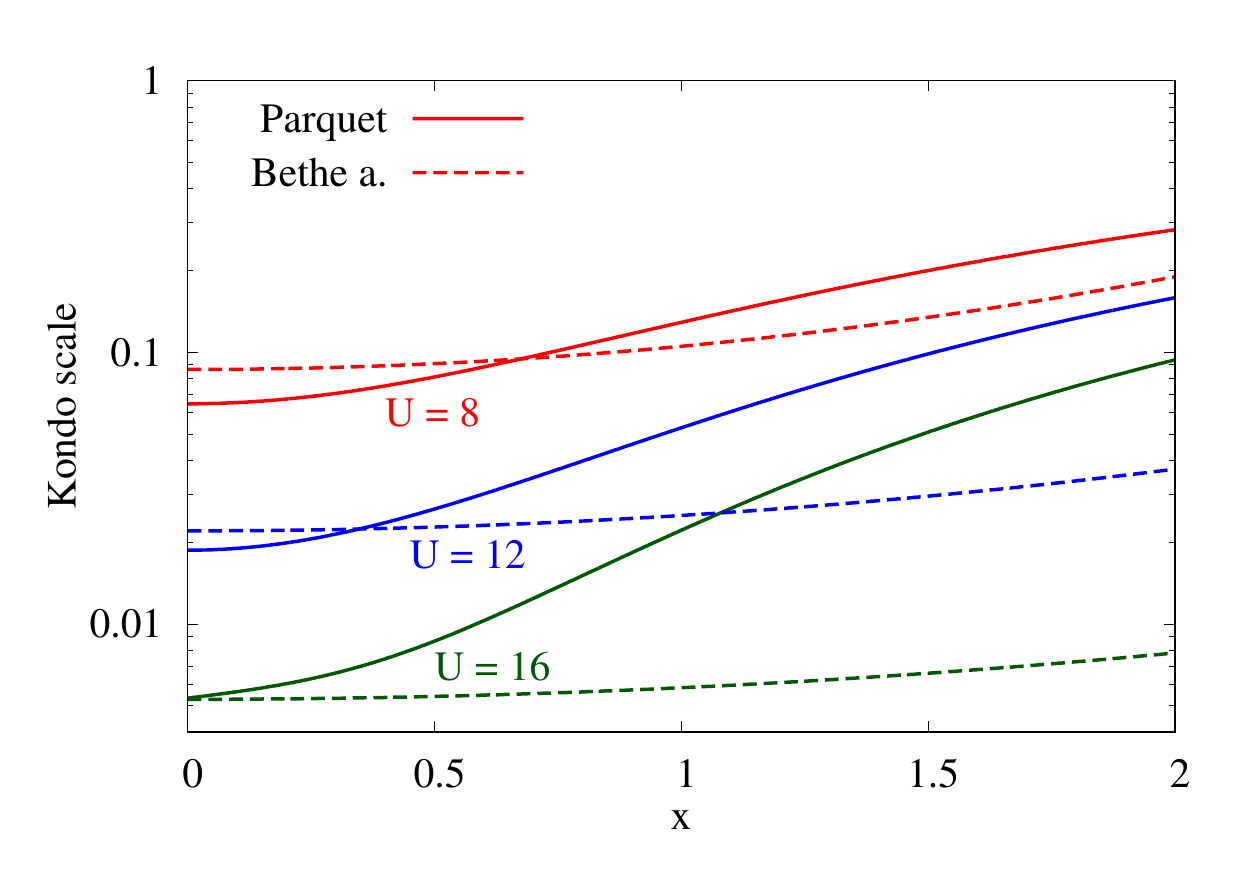}
\caption{(Color online) Kondo scale calculated from the denominator of the vertex function $K(\omega)$ at the Fermi energy $\omega=0$ as a function of the doping parameter $x$ compared for a few interaction strengths with the exact Bethe ansatz result.   \label{fig:Kondo-x}}
\end{figure}

Similarly we calculated and compared the Kondo scale as a function of the doping parameter $x$ for a few fixed interaction strengths in Fig.~\ref{fig:Kondo-x}. Kondo scale is no longer exponentially small, we are far from the critical point of the two-particle vertex and only a qualitative behavior of the Kondo scale as a function doping can be expected. The difference between our approximation  and the exact solution increases with doping  and stronger interaction away from half filling.  The simplest static approximation is insufficient to achieve a good quantitative agreement with the exact result there, since the farther from half filling the more relevant the dynamical fluctuations due to multiple electron-electron scatterings become. All fluctuations away from the Kondo regime are bounded, of the same order,  and the electron-hole scatterings lose on dominance. A dynamical approximation for the irreducible vertex $\Lambda$ is needed to achieve more accurate quantitative results. 

\subsection{Lattice model}

\begin{figure}
\includegraphics[width=7.5cm]{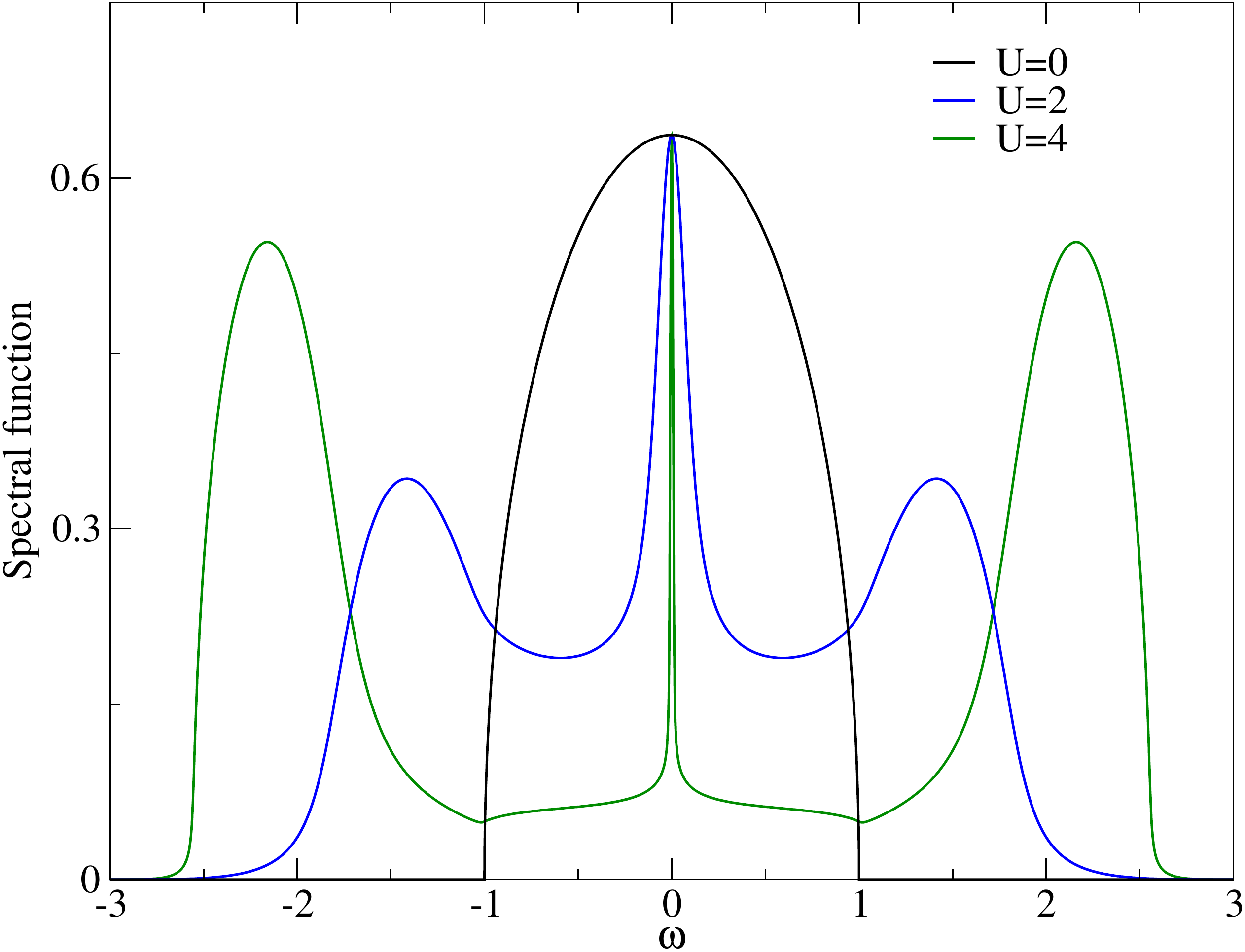}
\caption{(Color online) Spectral function in the mean-field approximation for the spectral self-energy of the Hubbard model at zero temperature with the semi-elliptic density of states at half-filling.   \label{fig:spectral_eliptic-symm}}
\end{figure}

The effective interaction is used to generate a mean-field approximation, that is, a local approximation to lattice models of interacting electrons. In the mean-field approximation we need to know only the density of states of the underlying lattice. We use the semi-elliptic density of states $\rho(\epsilon) = 2\sqrt{1 - \epsilon^{2}}/\pi$ corresponding to the infinite-dimensional Bethe lattice.  The halfwidth of the energy band is $w=1$. Using the equations from Sec.~\ref{sec:MFT} we calculate the spectral function obtained from the Green function renormalized with the spectral self-energy. The spectral function at zero temperature and half filling, $\mu=U/2$, is plotted in Fig.~\ref{fig:spectral_eliptic-symm} for several interaction strengths covering weak and intermediate coupling. We can see that the three-peak structure with a quasiparticle peak and Hubbard satellites develops rather soon already for a moderate interaction $U=2w$. This structure is amplified with increasing the  electron coupling. In the strong-coupling limit we see the same Kondo behavior as in the impurity model and no metal-insulator transition takes place, supposed that the low-frequency behavior of vertex $\Gamma(\omega)$ remains decisive.  We can also observe a non-analytic behavior in the region between the well formed central and satellite peaks due to the algebraic band edges of the density of states. The non-analyticity appears at the band edges of the noninteracting system. It is more pronounced when the renormalized density of states is small.      

\begin{figure}
\includegraphics[width=8.5cm]{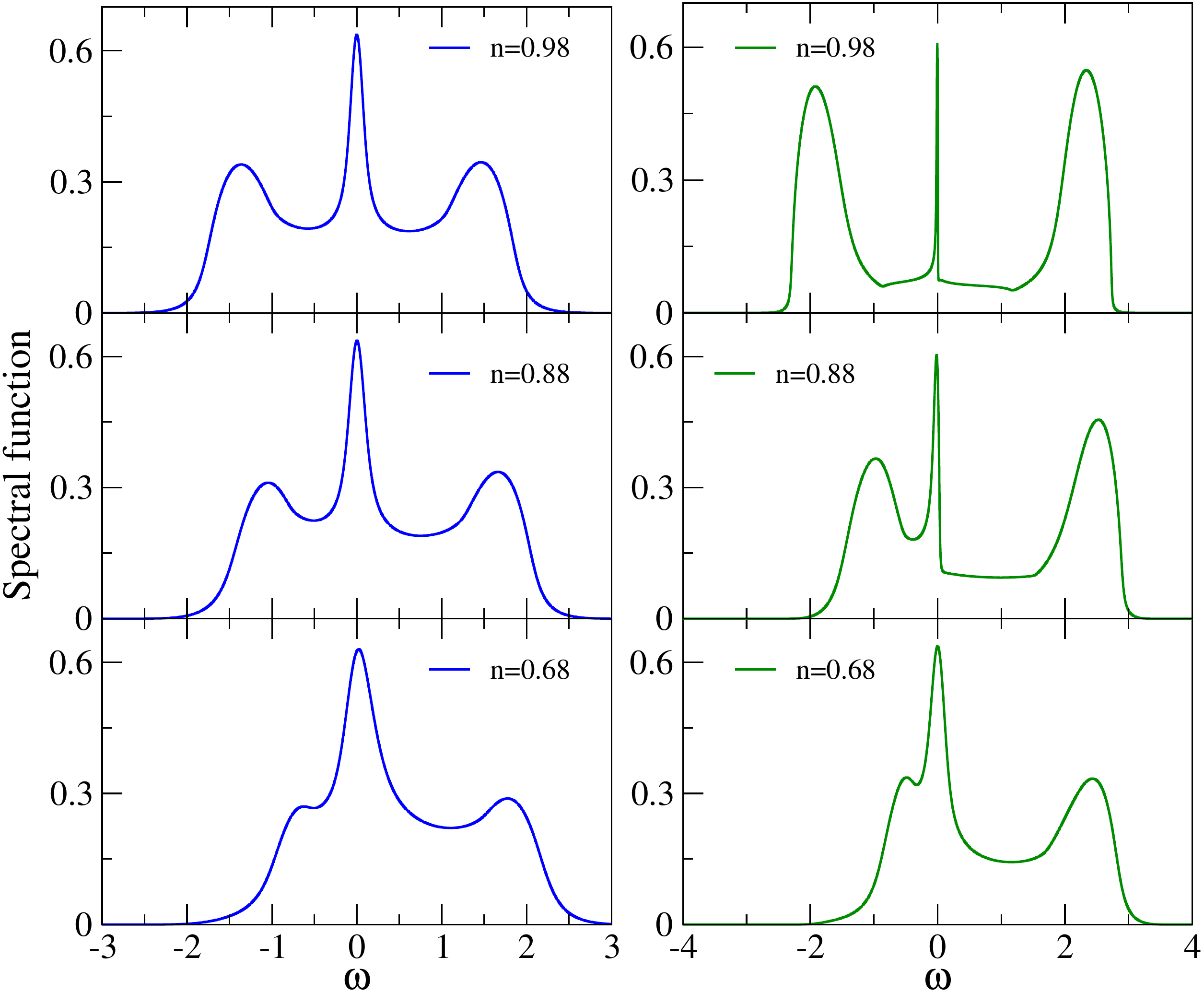}
\caption{(Color online) Spectral function in the mean-field approximation for the spectral self-energy  of the Hubbard model at zero temperature and off half-filling for $U=2$ (left pane) and $U=4$ (right pane).  \label{fig:spectral_eleptic-asymm}}
\end{figure}

We plotted in Fig.~\ref{fig:spectral_eleptic-asymm} development of the spectral function when doping the electron-hole symmetric state with holes $x<0$ for two interaction strengths. The behavior is very similar to the spectral function of the impurity model from Fig.~\ref{fig:spectral_sc}. The lower satellite band moves towards the central one and finally absorbs it (for $n<1$). The satellite peaks are higher for stronger interaction strengths. Notice also that the central and satellite peaks are always connected and the density of states between them never vanishes.  

\subsection{Magnetic susceptibilities}

The major asset of the two-particle approach based on the reduced parquet equations with the self-consistency via a linearized Ward identity is that the critical behavior of the two-particle vertex is transferred simultaneously to the spectral and response functions. In the impurity model it is the  exponentially small Kondo scale in the strong coupling. We discussed the dependence of the Kondo scale in the spectral function on doping in Sec.~\ref{sec:IMN} and the same should be seen also on the local magnetic susceptibility. 

\begin{figure}
\includegraphics[width=8cm]{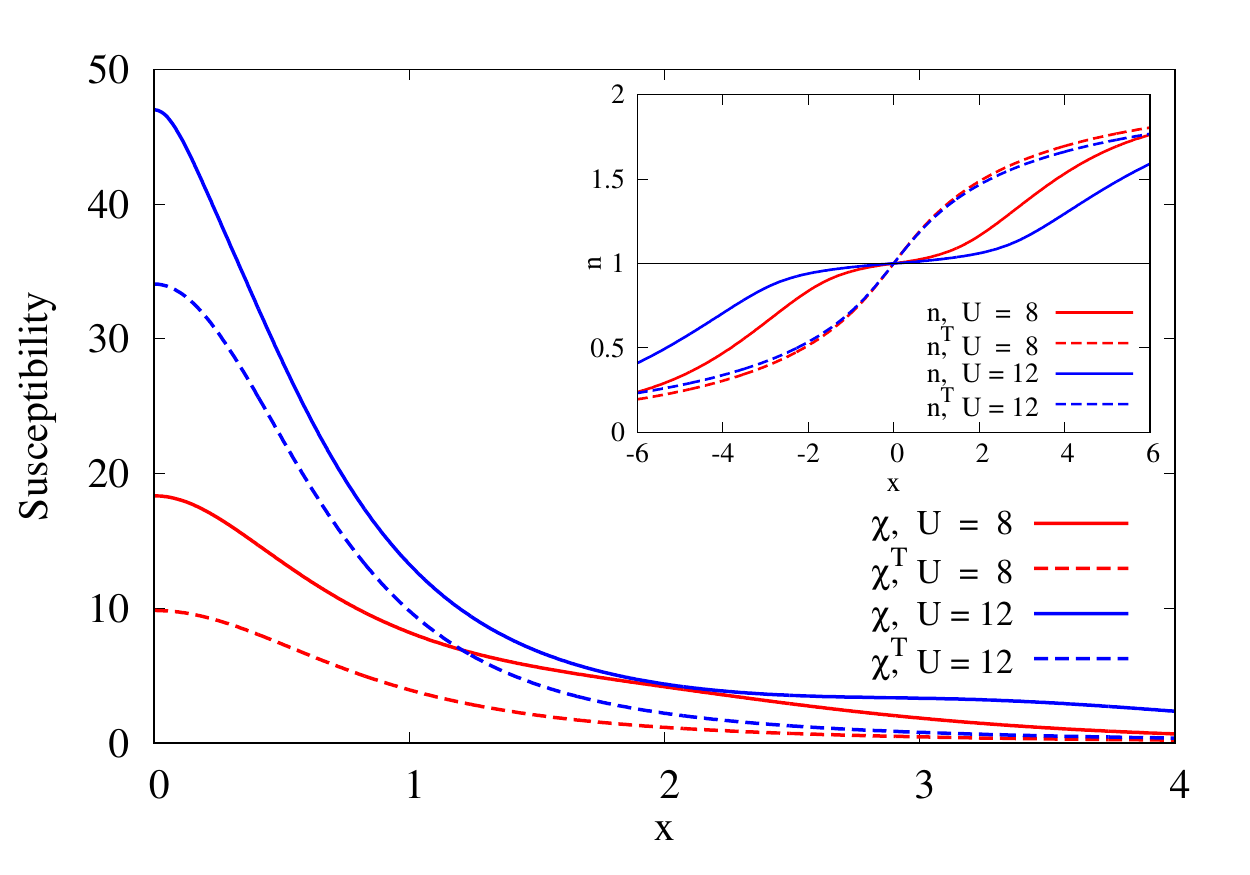}
\caption{(Color online) Dependence of the local magnetic susceptibilities of SIAM $\chi$ and $\chi^{T}$   on  the doping parameter $x$ for two interaction strengths. The inset shows the corresponding occupation numbers.    \label{fig:suscU_x}}
\end{figure}

We plotted in Fig.~\ref{fig:suscU_x} dependence of the zero-temperature magnetic susceptibility  of the impurity model (Lorentzian density of states) on the doping parameter $x$ for two interaction strengths in the strong-coupling limit. We plotted both thermodynamic $\chi^{T}$ and physical $\chi$ magnetic susceptibilities. A fast decay of their values with increasing doping is clearly demonstrated. The dependence is symmetric with respect to the electron-hole transformation $x\to -x$.  Both full and thermodynamic susceptibilities behave qualitatively in the same way. The inset shows the corresponding band fillings.         

\begin{figure}
\includegraphics[width=8cm]{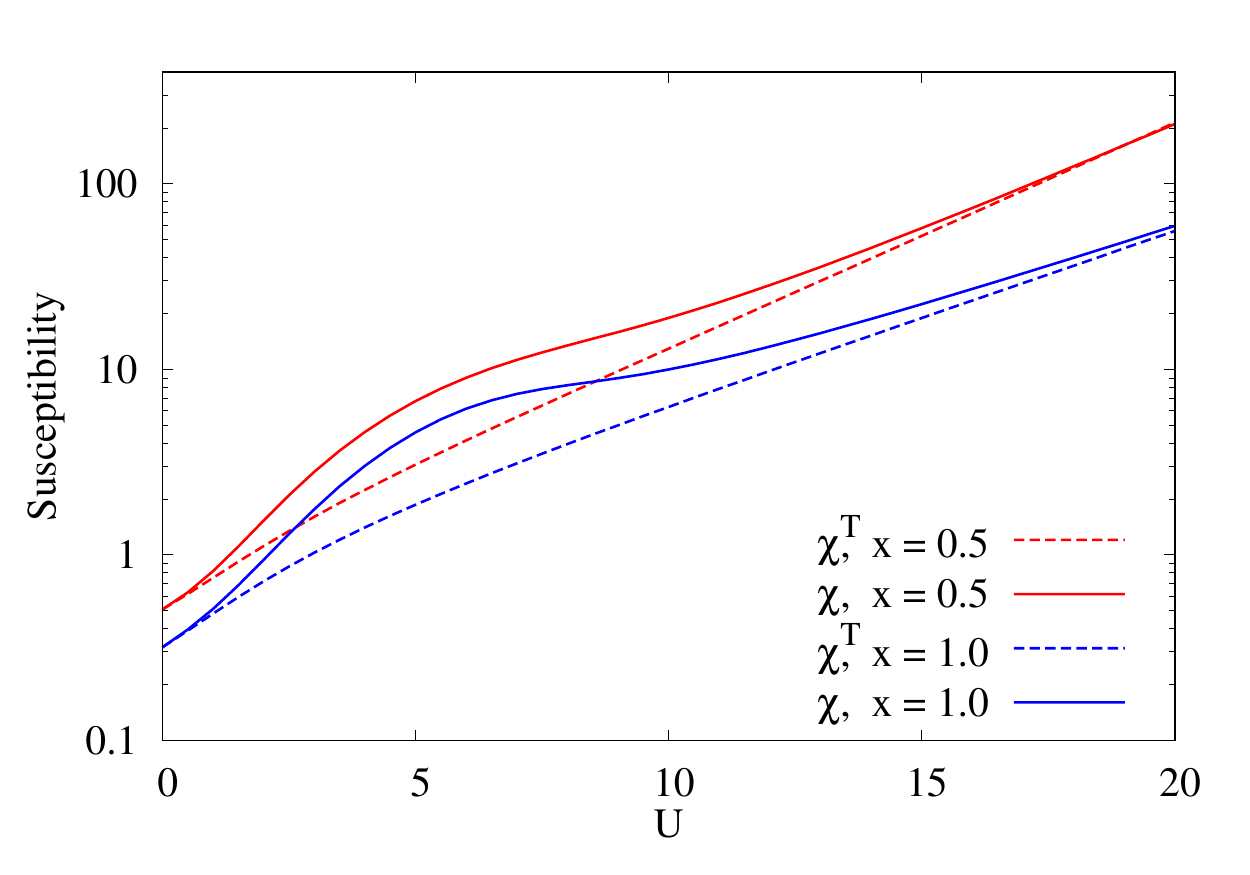}
\caption{(Color online) Local magnetic susceptibilities of SIAM $\chi$ and $\chi^{T}$  as a function of the interaction strength $U$ for fixed doping parameters $x$ in a logarithmic scale.   \label{fig:suscx_U}}
\end{figure}

If we perform the limit to strong coupling at a fixed doping parameter $x$ the system tends to reach the critical Kondo asymptotics with an exponentially small Kondo scale. Such a behavior is confirmed in the magnetic susceptibility where it exponentially grows with increasing the interaction strength, as demonstrated on Fig.~\ref{fig:suscx_U}. Both susceptibilities coincide in the strong coupling regime, but they differ in the noncritical region. The full susceptibility slows down its growth before it reaches its strong-coupling asymptotics. The thermodynamic susceptibility also bends its growth but less apparently.

\begin{figure}
\includegraphics[width=8cm]{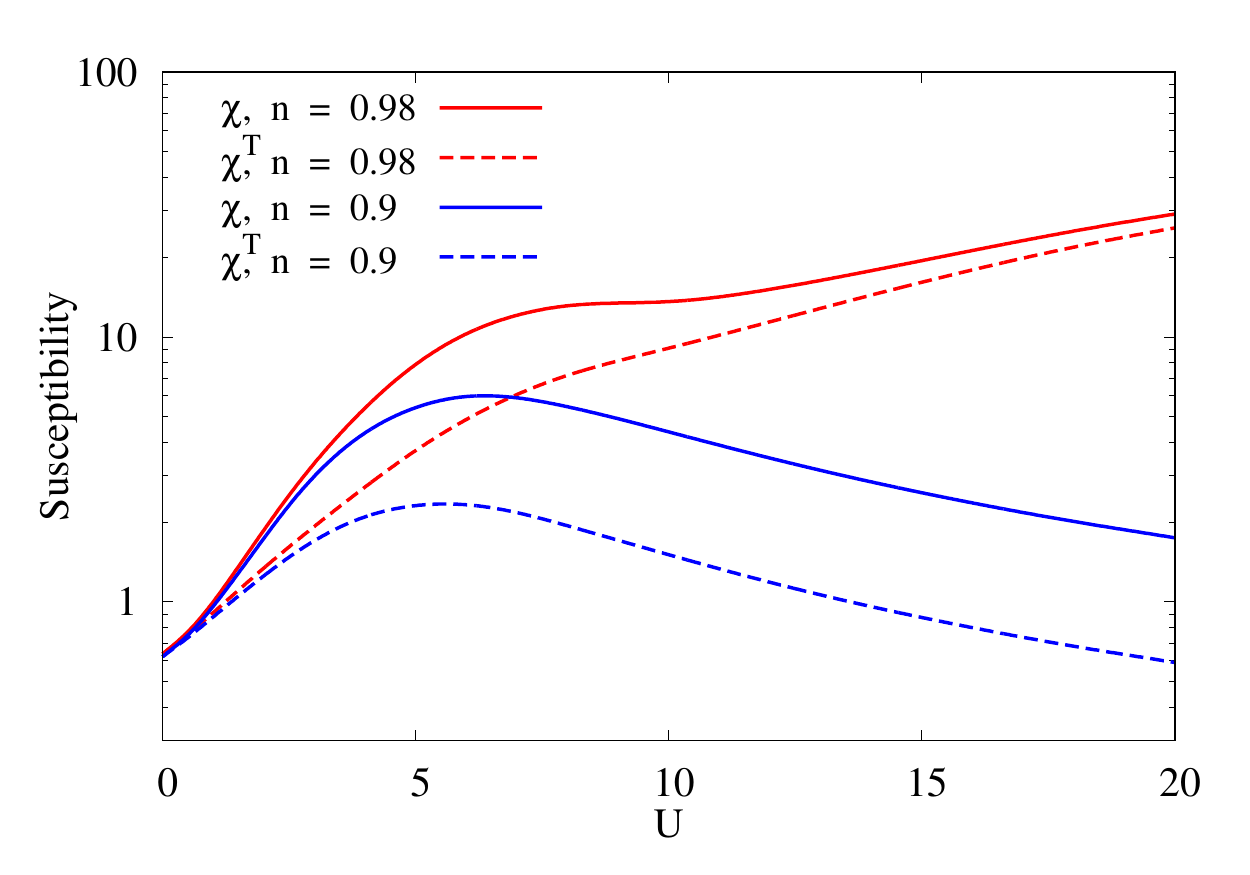}
\caption{(Color online) Local magnetic susceptibilities of SIAM $\chi$ and $\chi^{T}$  as a function of the interaction strength $U$ for fixed occupation numbers $n$ in a logarithmic scale.     \label{fig:suscn_U}}
\end{figure}

The fixed doping parameter $x$ does not mean fixed occupation number $n$ when going to the strong-coupling regime. It is hence interesting to plot dependence of the susceptibility on the interaction strength at a fixed $n$. Fig.~\ref{fig:suscn_U} shows dependence of the magnetic susceptibility for two fixed occupation numbers, one close to half filling and another farther away. We can see a big difference in shape. When close to half filling, the susceptibility grows exponentially in the strong coupling. While significantly away from half filling we can see that the magnetic susceptibility saturates at a finite interaction strength and starts do decrease, that is, the Kondo scale grows and the system moves away from the critical regime. It corresponds to broadening of the quasiparticle peak in the spectral function and its merger with one of the satellite Hubbard bands. As discussed in Sec.~\ref{sec:IMN} there is no exponentially small Kondo scale in the limit $U\to\infty$ for the fixed occupation number $n$.

\begin{figure}
\includegraphics[width=8cm]{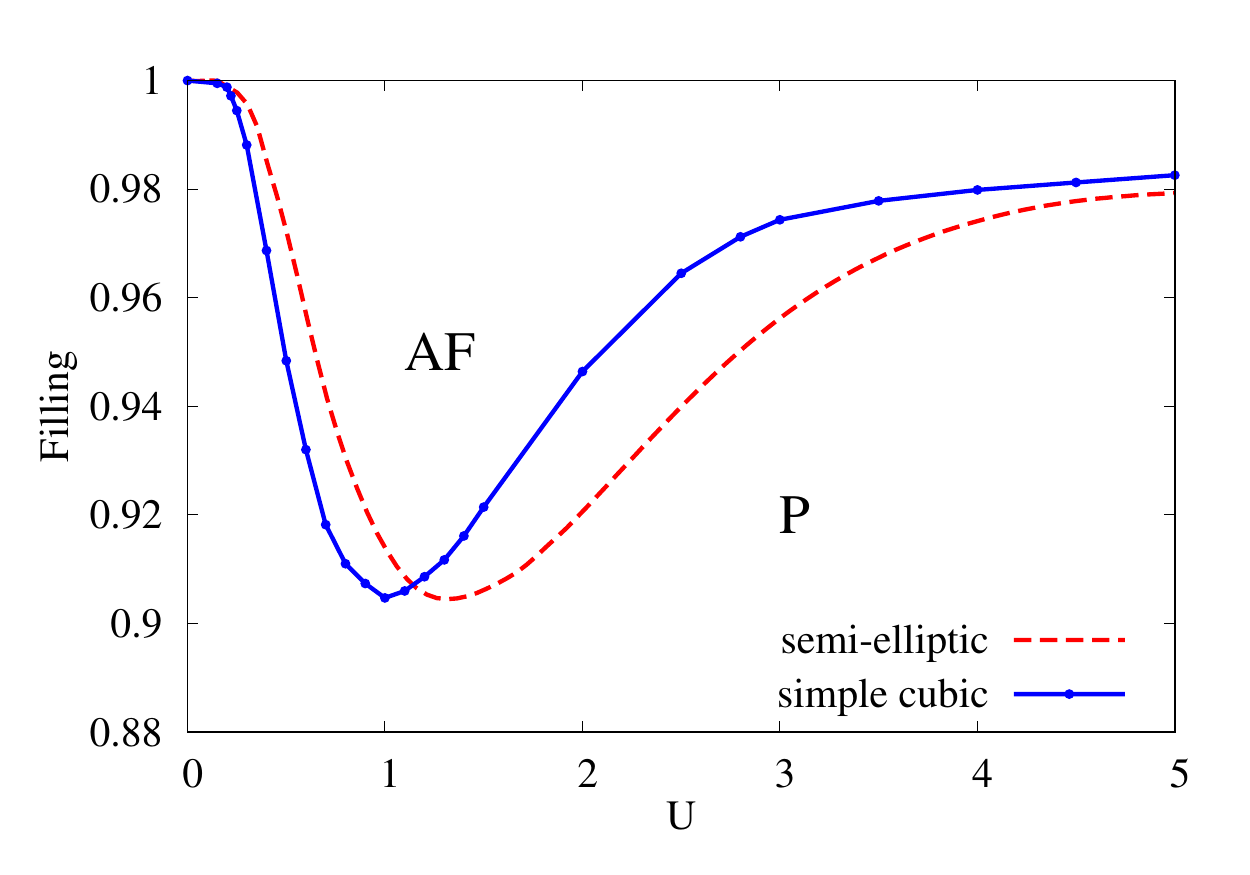}
\caption{(Color online) Mean-field phase diagram of the Hubbard model at zero temperature for the paramagnet-antiferromagnet  transition in $n-U$ plane.  Solid (blue) line for the cubic DOS, the dashed (red) line for the semi-elliptic DOS.    \label{fig:phase-AF}}
\end{figure}

The impurity model does not allow for a magnetic order. To apply the mean-field approximation also on extended lattice models one has to take into account magnetic ordering. Our approximation works best near the charge-symmetric situation where one expects antiferromagnetic ordering in homogeneous situations. We hence use the non-local thermodynamic susceptibility $\chi^{T}(0,\mathbf{Q})$ at zero temperature from Eq.~\eqref{eq:chiT-dynamic} to determine the boundary between the paramagnetic (P) and antiferromagnetic (AF) phases in the $n-U$ plane. We plotted in Fig.~\ref{fig:phase-AF} the phase boundaries for the simple cubic and semi-elliptic densities of states with half-bandwidth $w=1$.  One can see that the phase diagram has a universal structure with the strongest tendency towards the antiferromagnetic order in the intermediate coupling. A similar phase diagram was obtained in other approaches for the half-filled state of the infinite-dimensional Hubbard model in the $T-U$ plane.\cite{Jarrell:1992aa,Georges:1993aa} We can see that the phase boundary is not much affected by the particular form of the underlying density of states.     

\section{Conclusions}

 We extended in this paper the effective interaction approximation  on the electron-hole irreducible vertex in the reduced parquet equations from Ref.~\onlinecite{Janis:2016ac} to a static mean-field approximation for thermodynamic and spectral properties of strongly correlated electron systems. The main idea of the  mean-field approximation for the thermodynamic quantities is to replace the bare interaction $U$ of the Hartree approximation by a renormalized effective interaction $\Lambda$. The latter is then self-consistently determined from the reduced parquet equations and  the spurious phase transitions of the Hartree approximation are thereby suppressed. The one-electron propagators in the parquet equations are renormalized in the Hartree sense where the bare interaction is replaced with the effective one. Finally, a static correction to the thermodynamic self-energy is introduced away from half filling to maintain the exact electron-hole symmetry and to keep compressibility non-negative. A static, fully self-consistent thermodynamic mean-field approximation is thereby accomplished.

The static thermodynamic mean-field theory neglects completely the dynamical character of the interaction-induced fluctuations. They are derived from the dynamical vertex and the Schwinger-Dyson equation.  The  Schwinger-Dyson equation with the one-electron propagators renormalized with the thermodynamic self-energy is used to determine the spectral self-energy.  The full physical self-energy consists of a static term and the spectral self-energy. The spin-symmetric static part of the full self-energy must be determined from a self-consistent equation for the total particle density so that to keep compressibility non-negative.  We showed that the critical behavior of the thermodynamic mean-field approximation is qualitatively the same as that derived from the full dynamical self-energy.  As the thermodynamic mean-field theory is an extension of the Hartree approximation to strong coupling, the spectral self-energy from the Schwinger-Dyson equation with the mean-field propagators is then an extension of the second-order perturbation expansion around the  Hartree solution. 
         
We demonstrated equivalence of the critical behavior in the spectral and  thermodynamic functions on the single-impurity Anderson model where the critical behavior manifests itself via the Kondo effect near the half-filled impurity level. The qualitative Kondo behavior, being the linear dependence of the logarithm of the Kondo scale on the interaction strength, is reproduced in the effective-interaction approximation. Quantitative agreement with the numerically exact methods (NRG and QMC) decreases with doping the charge-symmetric state with holes or electrons as we recede from the Kondo regime. The static mean-field approximation remains qualitatively correct also outside of the critical regions of singularities in the Bethe-Salpeter equations with no critical fluctuations but it cannot be expected to give good quantitative results. A more advanced approximation with a dynamical irreducible vertex in the reduced parquet equations is needed.  

The one and two-particle functions from the thermodynamic mean-field approximation used to determine the physical self-energy  guarantee that the  critical behavior in both thermodynamic and spectral functions is qualitatively the same and leads to a consistent and reliable low-temperature critical behavior of correlated electron systems. Its universality, relative simplicity, and analytic control of the critical behavior make this approximate construction easily generalizable  to complex and realistic electron systems the phase diagram of which is presently achievable only with demanding and expensive numerical means.   

\section*{Acknowledgments}
 Research on this problem was supported by Grant No. 15-14259S of the Czech Science Foundation. We thank M. \v Zonda for providing us with NRG data.  


\end{document}